\documentclass{jfm}
\usepackage{graphicx}
\usepackage{epstopdf, epsfig}

\title{Vaporization frequency response to pressure oscillations: an approximate analytical solution for mixed injection regimes}

\author{Kwassi Anani\aff{1}
  \corresp{\email{kanani@univ-lome.tg}},
  R. Prud'homme\aff{2}
 \and M. N. Hounkonnou\aff{3}}

\affiliation{\aff{1}D\'epartement de Math\'ematiques, Universit\'e de Lom\'e,
Lom\'e, 02 BP 1515 Lom\'e, Togo
\aff{2}Jean Le Rond d'Alembert Institute, UMR 7190 - Pierre et Marie Curie University, 75252 Paris Cedex 05, Paris, France
\aff{3}International Chair in Mathematical Physics and Applications, Universit\'e d'Abomey-Calavi, 072 BP 050 Cotonou, Cotonou, B\'enin}

\begin{document}

\maketitle
\thispagestyle{empty}

\begin{abstract}
This work is devoted to a theoretical analysis of mass frequency response to pressure oscillations of a spray of repetitively injected drops into a combustion chamber. A single stationary spherical droplet continuously fed with the same liquid fuel so that its volume remains constant despite the evaporation, the so-called 'mean droplet' in the Heidmann analogy, represents this vaporizing spray. The feeding is realized with a liquid-liquid heat transfer coefficient by using a source point placed at the mean droplet centre, in such a way that only radial thermal convection and conduction effects are allowed inside the droplet during the process. This feeding procedure is now viewed as a proper boundary condition that is a mixed or a generalized feeding regime controlling the liquid fuel injection process into the combustion chamber. Drawing upon a linear analysis based on the Rayleigh criterion, the evaporating mass response factor is evaluated. Effects due to the variation of the heat transfer coefficient and that of the process characteristic times are analysed. An abrupt increase appears in the response when a fuel thermodynamic coefficient approaches a particular value. 
\end{abstract}

\begin{center}
\small\today  
\end{center}

\section{Introduction}
Combustion instabilities still nowadays a challenging area in combustion research though their modelling and control have been investigated in many published works by various research teams during the past decades. Combustion instabilities result from the coupling between acoustic waves and combustion. In confined devices, the coupling between acoustic field and heat or mass release at certain frequency levels may lead to engine failure or other catastrophic consequences \citep{Nair14}. On the contrary, new blends of fuels can be engineered to undergo preferential instabilities leading to homogeneous combustion with higher efficiency \citep{Candel13}. The present paper aims at contributing to the linear analysis of subcritical combustion instabilities by analytical approaches based on the mean spherical droplet configuration as in \citet{Anani18}. In the following section, a brief description is made of the unperturbed state corresponding to the vaporization of the continuously fed spherical droplet in a stable environment. In \S\ref{sec3}, the linear analysis for harmonic perturbations in pressure is performed and a double confluent Heun equation \citep[see][]{Slavyanov00} is derived from the energy equation of the liquid phase. An approximate analytical expression of the temperature profile inside the mean droplet is then obtained for the generalized or mixed injection regime and the mass response factor is defined. Results are discussed in \S\ref{sec4}. Throughout the discussion, comparisons are made with results of certain models in the literature that account for the actual changing volume due to vaporization of individual injected droplets in the spray. Finally, key results are recalled in the conclusions.

\section{Stabilized state description}\label{sec2}

\subsection{General assumptions}
Individual spherical fuel droplets are repetitively injected into a subcritical combustion chamber. The distance between the droplets is supposed large enough, so that no interaction occurs between the droplets or between the droplets and the wall. Assuming velocity-stabilized hypotheses as in \citet{Heidmann66}, the liquid fuel vaporizing spray is represented by an idealized physical configuration of a mean spherical droplet at rest in the combustion chamber. The mean droplet, placed at a specified location in the combustion chamber (pressure anti-node and velocity node), is supposed to summarize the frequency response of individual drops in the spray. The vaporizing mean droplet has a constant average radius $\bar r_S$ since its instantaneous evaporating mass $\dot M$ is continuously restored with an average mass flow rate $\overline{\dot M}$ of the same fluid by using a point source placed at the centre. The choice of the Arithmetic Mean Diameter configuration is motivated by the analytical approach of the problem since it leads for the mean droplet to conservation equations with fixed boundary conditions. From now on, all barred quantities indicate mean values corresponding to the stabilized state whereas all primed quantities will denote perturbed quantities i.e. $ x' = (x - \bar x)/\bar x$.

\begin{figure}[!h]
  \centerline{\includegraphics{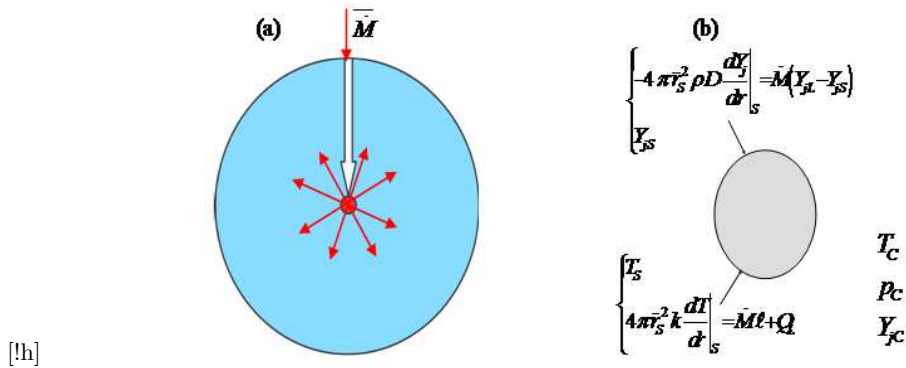}}
  \caption{(a) The mean vaporizing droplet, continuously fed by a point source placed at its centre. (b) Boundary conditions for the supplied droplet.}
\label{fig1}
\end{figure}

The local feeding rate $\overline{\dot M}$ is distributed throughout the droplet (see figure \ref{fig1}(a)) in such a way that, except for the radial thermal convection effect from the droplet centre to its evaporation surface, other convective transport or liquid recirculation phenomenon within the droplet are negligible. The spherical symmetry of the mean droplet is maintained at every moment during the process, and the thermal dilatation of the liquid is negligible so that the density $\rho_L$, the specific heat $c_L$ and the thermal conductivity $ k_L $ of the droplet will be treated as constant. At the mean droplet centre, a generalized or mixed boundary condition is considered, that is the liquid fuel is injected with a positive heat transfer coefficient $ h $. The two extreme cases of this injection process are the adiabatic feeding regime ($ h=0 $) where zero temperature gradient is assumed at the droplet centre, and the isothermal feeding regime ($ h=\infty $) where the droplet centre is kept at a constant temperature $\overline {T}_S $. The latter is the mean value of the spatially uniform but time-varying temperature ${T}_S $ of saturated vapour at the stabilized droplet surface. 

In the immediate vicinity of the droplet surface, the gas phase is made up of stoichiometric reaction products evolving in a quasi-steady state. Equilibrium conditions at the droplet/gas interface are assumed for the stabilized state and there is no gas diffusion into the droplet. Far from the mean evaporating droplet, the ambient environment inside the chamber is at constant subcritical temperature $ T_C $ and pressure $p_C$. The boundary conditions at the supplied droplet surface are shown in figure \ref{fig1}(b). Subscripts $ L $  and  $ l $ refer to liquid phase, whereas subscripts $ S $  and $ C $ respectively indicate the droplet surface and the conditions for the combustion chamber far from the droplet. The heat flux transferred to the liquid is designated by $ Q_L $ and the binary diffusion coefficient of fuel vapour in air is denoted by $ D $. The density and the thermal conductivity of the gas mixture around the droplet surface are respectively designated by $ \rho $ and $ k $. The mass fraction of species $ j $ being denoted by $ Y_j $, the gaseous mixture near the surface is composed of fuel species $ j=F $ and of combustion products diluted species $ j=A $ proceeding from the flame front at infinity. For reason of simplicity, we have considered a mono-component droplet with only fuel species, that is   $ Y_{FL}=1 $ and $ Y_{AL}=0 $. 
\subsection{Characteristic times}
The residence time of the continuously fed droplet can be equated with the mean lifetime of an individual vaporizing droplet in the spray. This time replaces the notion of the free droplet lifetime in the present situation of constant volume and is identified to the ratio  $ \bar \tau _v  = \overline M /\overline {\dot M} $, where $ \overline M $ represents the mean value of the actual mass $ M $ of the supplied droplet and $ \overline {\dot M} $ is the stationary feeding rate. The transfer time by thermal diffusion process is defined as $\bar \tau _T  = \bar r_S^2 /\kappa _L $, where $\kappa _L  = k_L /(\rho _L c_L) $ is the thermal diffusivity of the liquid and $ \bar r_S $ the constant average radius. It is then convenient to use the timescale ratio $ \theta  = 9\bar \tau _v /\bar \tau _T  = \bar \tau _v /\tilde \tau _T  $, which will be called from now on the thermal exchange ratio or more briefly the exchange ratio, as it is of the same order of magnitude as $ 1/\Pen_L $, $\Pen_L $ being the P\'eclet number of the liquid phase. The coefficient 9 is kept for comparison purposes with results obtained in \citet{Anani18}. During the vaporization, intrinsic or external pressure-related oscillations can cause departure from stabilized-state conditions. The frequency of such ambient pressure oscillations is a major characteristic time of the process. In the case of small harmonic perturbations in pressure, a linear analysis can be performed. The frequency of the harmonic oscillations in ambient pressure will be denoted by $\omega $. In order to provide a parameter depending on the residence time $ \bar \tau _v $, which may be used to characterize the frequency response for classical fuels, a reduced frequency $ u $ defined as $ u=3 \omega \bar \tau _v $ will be considered.  
\subsection{Unperturbed state equations}
The mass balance of the mean droplet is: 
\begin{equation}
\frac{dM}{dt} = \overline{\dot M}-\dot M,
\label{eq1}
\end{equation}
with $ \overline {\dot M} $ denoting the stationary flow of injection and $ \dot M $ the instantaneous flow of evaporation. In a stabilized state, one has: $\dot M \equiv \overline {\dot M}$, $dM/dt = 0$ and $M = \overline M $. The amount of heat $Q_L$ penetrating into the droplet is expressed as:
\begin{equation}
Q_L  = Q - \dot M\ell = 4\upi \bar r_S^2 k_L {\partial T_l}/{\partial r}\quad \mbox{on\ }\quad r=\bar r_S,
\label{eq2}
\end{equation}
where $T_l  \equiv T_l (r,\,t)$ is the temperature value at radial coordinate $r$ and time $t$ inside the mean droplet. The external gas heat flux is denoted by $Q$  and $\ell$ designates the latent heat of vaporization per unit mass of the liquid. Equation (\ref{eq2}) assures the coupling of the gas and the liquid phase solutions at the mean droplet surface. The formulation of the energy conservation equation includes both radial thermal convection and conduction data. In these conditions, the internal temperature $T_l$  satisfies the following equation: 	
\begin{equation}
\rho _L c_L \frac{{\partial T_l }}{{\partial t}} + \rho _L c_L v_r \frac{{\partial T_l }}{{\partial r}} - \frac{{k_L }}{r}\frac{{\partial ^2 \left( {rT_l } \right)}}{{\partial r^2 }} = 0,
\label{eq3}
\end{equation}                                      
where $v_r$ is the central injection velocity expressed as $v_r  = \overline {\dot M} /4\upi \rho _L r^2 $ and  $0 < r < \bar r_S $. This equation is solved, subject to the mixed boundary condition at the droplet centre and to the Dirichlet boundary condition at the surface:
\begin{equation}
\left. \begin{array}{ll}  
\displaystyle{\frac{{\partial T_l }}{{\partial r}}}   = \frac{h}{{\bar r_S }}\left( {T_l - {\overline {T}}_S } \right)
  \quad \mbox{on\ }\quad r=0,\\[8pt]
\displaystyle  T_l = T_S
  \quad \mbox{on\ }\quad r=\bar r_S.
 \end{array}\right\}
  \label{eq4}
\end{equation}

Assuming quasi-steady hypotheses, the droplet surface is in local evaporation equilibrium and the instantaneous mass vaporization rate can be calculated as:
\begin{equation}
\dot M = 2\upi \rho D r_S Sh^ *\ln({1 + B_M }) = 4\upi\frac{k}{{c_p }}r_S Nu^ *\ln ( {1 + B_T }),
\label{eq5}
\end{equation}
where $B_M  = (Y_{FS}  - Y_{FC} )/(1 - Y_{FS} )$ and $B_T  = c_p ( {T_C  - T_S } )/(\ell  + {Q_L }{\dot M}^{ - 1} )$ are the well-known Spalding mass and heat transfer numbers, and  $c_p$ the specific heat capacity of fuel vapour at constant pressure. As mentioned above, parameters $\rho$, $k$ , and $D$ are the density, the thermal conductivity and the binary diffusion coefficient of the mixture of vapour and ambient gas. The Sherwood and Nusselt numbers $Sh^ *$ and $Nu^ *$ were provided by \citet{Abramzon89} in their extended film model. At the droplet surface, the saturated vapour pressure can be expressed as $p_{sat} \left( {T_S } \right) = \exp \,\left( {a - b/(T_S  - c)} \right)$, where $a$, $b$ and $c$ are some coefficients related to the fuel thermophysical properties. The pressure $p_{sat}$  and the mole fraction $X_{FS}$ of fuel species are connected by the relation $ p\,X_{FS}  = p_{sat} ( {T_S } )$, where $p=p_C$ denotes the ambient pressure. If the molecular weight of species $j$(=$A$ or $F$) is denoted by ${\mathcal{M}}_j$, then the mass fraction $Y_{FS}$ of the vapour at the droplet surface can be written as a function of the mole fraction $X_{FS}$ as $Y_{FS} =\mathcal{M}_F X_{FS}/({\mathcal{M}}_F X_{FS}  + {\mathcal{M}}_A X_{AS})$. Since concentrations and temperature values are varying in the gas phase, the averaged properties can be evaluated at some reference concentration $\overline Y_j  = Y_{jS}  + A_r ( {Y_{jC}  - Y_{jS} } )$ and temperature $\overline T = T_S  + A_r( {T_C  - T_S })$ with $A_r=\mathrm{1/3}$. Both $Sh^*$ and $Nu^*$ are assumed equal to 2 and the Lewis number $Le = k/\rho Dc_p$ equal to 1.
\section{Linear analysis for small perturbations}\label{sec3}
\subsection{Linear analysis of the liquid-phase equations}
Splitting up the flow variables into steady and unsteady parts can be realized by writing $\Delta f = f - \bar f$, where $f$ is a flow parameter, $\bar f$ is its mean value, $\Delta f$ is the absolute perturbation, and $f^{\prime} = \Delta f/\bar f$ is the corresponding relative perturbation. The heat flow at the surface (equation (\ref{eq2})) is then given by:
\begin{equation}
{4\upi} \bar r_S^2 k_L \overline {T}_S \partial T_l^{\prime} /\partial r = Q_L  = Q_L  - \bar Q_L= \Delta Q_L \quad \mbox{on\ }\quad r=\bar r_S, 
\label{eq7}
\end{equation}
as $\bar Q_L=0$. For the perturbed temperature $T_l^{\prime} (r,t) = (T_l (r,t) - \overline T_l (r,t))/\overline T_l (r,t)$, the energy conservation equation (\ref{eq3}), can be rewritten as:
\begin{equation}
\frac{\partial (rT_l^{\prime} )}{\partial t} + \kappa _L \left( \frac{3 \bar r_S}{\theta  r}\frac{\partial T_l^{\prime}}{\partial r} - \frac{\partial ^2 ( {r T_l^{\prime}} )}{\partial r^2 }\right) = 0, 
\label{eq8}
\end{equation}
where $\theta  = \bar \tau _v /\tilde \tau _T$ is the thermal exchange ratio (see section \ref{sec2}). The perturbed boundary conditions in the mixed feeding regime are deduced from equation (\ref{eq4}) as follows: 
\begin{equation}
\left. \begin{array}{ll}  
\displaystyle{\frac{{\partial T_l^{\prime} }}{{\partial r}}}   = \frac{h}{{\bar r_S }}{T_l^{\prime}}
  \quad \mbox{on\ }\quad r=0,\\[8pt]
\displaystyle  T_l^{\prime} = T_S^{\prime}
  \quad \mbox{on\ }\quad r= \bar r_S.
 \end{array}\right\}
  \label{eq9}
\end{equation}
Introducing now small harmonic perturbations of frequency $\omega$ in the form of $
f^{\prime} = \hat f ( r )\mathrm{e}^{\mathrm{i}\omega t}$, the ambient pressure $p_C$ becomes $p^{\prime} = \hat p_C \mathrm{e}^{\mathrm{i}\omega t}$, while the temperature is expressed as $T_l^{\prime}  = \hat T_l ( r )\mathrm{e}^{\mathrm{i}\omega t}$, and the heat transferred into the droplet as $\Delta Q_L  = \Delta \hat Q_L (r)\mathrm{e}^{\mathrm{i} \omega t}$. Equation (\ref{eq8}) is then transformed into: 
\begin{equation}
\mathrm{i} r^2 \omega \hat T_l  + \frac{3\kappa _L \bar r_S}{\theta }\frac{d\hat T_l}{dr} - \kappa _L r\frac{d^2 ( {r \hat T_l } )}{dr^2 } = 0, 
\label{eq10}
\end{equation}
or equivalently into: 
\begin{equation}
\mathrm{i} \omega \bar \tau _T \xi \hat T_l  + \frac{{1}}{3\theta  \xi}\frac{d\hat T_l}{d\xi} - \frac{d^2 ( {\xi \hat T_l } )}{d\xi ^2 } = 0, 
\label{eq11}
\end{equation}
where $\hat T_l$ is taken as a function of the reduced radius variable $\xi=r/\bar r_S$ ($0 < \xi  < 1$). The boundary conditions in the generalized feeding regime (equations (\ref{eq9})) can then be written in connection with $\xi$ as:
\begin{equation}
\left. \begin{array}{ll}  
\displaystyle \frac{d\hat T_l}{d\xi}   = \frac{h}{{\bar r_S }}{\hat T_0 }
  \quad \mbox{on\ }\quad \xi=0,\\[8pt]
\displaystyle  \hat T_l = \hat T_S
  \quad \mbox{on\ }\quad \xi= 1,
 \end{array}\right\}
  \label{eq12}
\end{equation}
where $\hat T_0$ depends on the initial temperature of the injected liquid fuel.

We now consider the complex number $\bar s_0  = ( {1 - \mathrm{i}} )({\omega /2\kappa _L })^{1/2}$, conjugate of  $ s_0  = ( {1 + \mathrm{i}} )({\omega /2\kappa _L })^{1/2}$, $s_0$  and $-s_0$  being the roots of the characteristic equation $\mathrm{i}\omega  - \kappa _L s^2  = 0$ obtained from equation (\ref{eq10}), when neglecting the convective term $(3\kappa _L \bar r_S /\theta )d\hat T_l /dr$. For any given value of the heat transfer coefficient $h>0$, a solution of equation (\ref{eq11}) subject to conditions (\ref{eq12}) can be sought in the form of $\xi \hat T_l (\xi ) = \mathrm{J}(\xi )\lbrace{1 - \cos [ {\bar s_0  \bar r_S \xi \exp (\mathrm{i}\arctan h)} ]} \rbrace$, with $\exp (\mathrm{i}\arctan h) = (\mathrm{i}h + 1)/({h^2  + 1})^{1/2}$, and $\mathrm{J}$ referring to a function to be determined. From the second-order truncated expansions of sine and cosine functions that are $\sin ( {S_0 \xi } ) \approx S_0 \xi$ and $\cos ( {S_0 \xi } ) \approx 1 - ( {S_0 \xi } )^2 /2$ with $S_0  = \bar s_0  \bar r_S\exp (\mathrm{i}\arctan h)$, it is deduced that the function $\xi \mathrm{J}$ approximately verifies the following double confluent Heun equation:
\begin{equation}
\frac{\xi ^2 {d^2 (\xi \mathrm{J})}}{{d\xi ^2 }} + \left( {2\xi  - \frac{3}{\theta}}\right)\frac{d(\xi \mathrm{J})}{d\xi} - 2\bar s_0^2 \bar r^2 _S \frac{h(\mathrm{i} - h)}{h^2  + 1}\xi ^2 (\xi \mathrm{J}) = 0. 
\label{eq13}
\end{equation}
By using Maple notation, a solution of equation (\ref{eq13}) can be expressed as: $\mathrm{J}(\xi) = C_0 \exp [ - 3(\theta \xi )^{ - 1}]{\mathrm{HeunD}}( {x_1 ,x_2 ,x_3 ,x_4 ,x})/\xi ^{5/2}$, where $C_0$ is an arbitrary constant and ${\mathrm{HeunD}}( {x_1 ,x_2 ,x_3 ,x_4 ,x})$ is the double confluent Heun function with its corresponding four parameters: $x_1  = 0$, $x_2  =  - [\theta ^2 (h^2  + 1) - 9 - 9h^2  - 24uh(\mathrm{i}h + 1)\theta ]/4\theta ^2 (h^2  + 1)$, $x_3  =  - [9 + (9 - 24\mathrm{i}u\theta )h^2  - 24hu\theta ]/2\theta ^2 (h^2  + 1)$ and $x_4  =  - [ - \theta ^2 (h^2  + 1) - 9 - 9h^2  - 24uh(\mathrm{i}h + 1)\theta ]/4\theta ^2 (h^2  + 1)$ and the variable $x=(\xi ^2  - 1)/(\xi ^2  + 1)$. We recall that the quantity $u = 3\omega \bar \tau _v$ is the ambient pressure frequency defined in the precedent section. Finally, the condition $\hat T_l (1) = \hat T_S$ at the mean droplet surface leads to an approximate analytical solution expressed as:
\begin{eqnarray}
\hat T_l (\xi ) = \frac{\hat T_S \lbrace{1 - \cos [ {\bar s_0  \bar r_S \xi \exp (\mathrm{i}\arctan h)} ]} \rbrace}{{\lbrace{1 - \cos [ {\bar s_0  \bar r_S \exp (\mathrm{i}\arctan h)} ]} \rbrace\xi ^{5/2} }}\exp \left[\frac{3}{2\theta}\left(1 - \frac{1}{\xi}\right)\right]\nonumber\\
\mbox{}\times {\mathrm{HeunD}}\left(x_1 ,x_2 ,x_3 ,x_4 ,\frac{\xi ^2  - 1}{\xi ^2  + 1}\right).
\label{eq14}
\end{eqnarray}
The above approximate analytical solution presents an essential discontinuity at    $\xi  = 0$ since, once $h>0$, the temperature gradient is not null at the droplet centre. Now, the calculation of the mass response factor only includes regularity conditions at the droplet surface $\xi  = 1$ and these conditions are well verified by this approximate solution. Thus, the flow condition at the droplet surface equation (\ref{eq7}) can be rewritten as $4\upi \bar r_S k_L \bar T_S \frac{d\hat T_l}{d\xi}(1) = \Delta \hat Q_L $, and then be applied to the solution (\ref{eq14}). That leads to:
\begin{equation}
\Delta \hat Q_L  =  - 4\pi\bar r_S k_L \overline T_S \hat T_S E( {u,\theta ,h } ), 
\label{eq15}
\end{equation}
where $E$ is expressed in function of $u$, $\theta$ and $h$ as:
\begin{equation}
E({u,\theta ,h}) = \frac{\bar s_0 \bar r_S \exp (\mathrm{i}\arctan h)\sin [\bar s_0 \bar r_S {\exp (\mathrm{i}\arctan h)}]}{\cos [ {\bar s_0  \bar r_S \exp (\mathrm{i}\arctan h)}]-1} - \frac{3}{2\theta} + \frac{5}{2}, 
\label{eq16}
\end{equation}
with $\bar s_0 \bar r_S = (1 - \mathrm{i})(3u/2\theta )^{1/2} $, $u = 3\omega \bar \tau _v $ and $\theta = \bar \tau _v /\tilde \tau _T $.
\subsection{Gas-phase linearized equations}
The linearized equations for the liquid/gas interface initially presented in \citet{Prudhomme10} are here briefly recalled. Introducing harmonic perturbation of the form $ f^\prime = \hat f\mathrm{e}^{\mathrm{i}\omega t} $, the ambient pressure is given by $ p^\prime = \hat p_C \mathrm{e}^{\mathrm{i}\omega t} $, and the mass flow rate by ${\dot M}^\prime = \hat {\dot M}\mathrm{e}^{\mathrm{i}\omega t}$. Consequently, the equations of the gas phase (see section \ref{sec2}) imply: 
\begin{equation}
\hat {\dot M} = \alpha\frac{\mathrm{i}u}{1 + \mathrm{i}u}( \bar b\hat T_S  - \hat p_C ),
\label{eq17}
\end{equation}
and:
\begin{equation}
\Delta \hat Q_L  = \overline {\dot M} \bar \ell ( \bar a\hat p_C  - \mu \hat T_S  ),
\label{eq18}
\end{equation}
where $u = 3\omega \bar \tau _v$ and $\Delta Q_L  = \Delta \hat Q_L \mathrm{e}^{\mathrm{i}\omega t}$. The coefficients involve in these equations are: 
\begin{equation}
\left. \begin{array}{ll}  
\displaystyle \bar a = \frac{\overline T_C }{\overline T_C  - \overline T_S}\frac{\gamma - 1}{\gamma}+ \varphi,\quad \bar b = \frac{\overline T_S}{(\overline T_S  - c)^2}b,\quad \mu = \frac{\overline T_S}{\overline T_C  - \overline T_S} - \frac{2c}{\overline T_S  - c} + \bar b\varphi,\\[8pt]
\displaystyle  \alpha  = \frac{\overline B_M}{(1 + \overline B_M)\ln (1 + \overline B_M )}\varphi; \quad  \varphi  = \frac{\overline Y_{AC}\overline Y_{FS}}{\overline Y_{AS} (\overline Y_{FS}  - \overline Y_{FC})}\frac{\mathcal{M}_F }{{\mathcal{M}}_F \overline X_{FS}  + {\mathcal{M}}_A \overline X_{AS}}.
 \end{array}\right\}
  \label{eq18b}
\end{equation}
The parameter $\gamma$ stands for the constant isentropic coefficient and the latent heat of vaporization $\ell$ per unit mass of the liquid is given by: $\ell  = b RT_S^2 /\mathcal{M}_F(T_S  - c)^2$, where $R$ denotes the universal gas constant.
\subsection{Mass response factor}

According to the Rayleigh criterion for sinusoidal oscillations which are uniform over a finite volume, the response factor is defined as $N=(\left| {\hat q} \right|/\left| {\hat p} \right|)\cos \phi$, where $\left| {\hat q} \right|$ and $\left| {\hat p} \right|$ are the moduli of mass release $q$ and pressure $p$ and $\phi$ is the phase difference between $q^\prime$ and  $p^\prime$. Therefore, a reduced mass response factor can be defined as the real part of the transfer function $Z =\hat {\dot M}/( \alpha \hat p_C  )$. By using equations (\ref{eq15})-(\ref{eq18}), $Z$ is deduced in function of $u$, $\theta$ and $h$ as: 
\begin{equation}
Z( {u,\theta ,h}) = \frac{\mathrm{i}u}{1 + \mathrm{i}u}\frac{A + \theta E( {u,\theta ,h})}{B - \theta E( {u,\theta ,h})},
\label{eq20}
\end{equation}
where $A = 3 (\bar a\bar b - \mu )/ \lambda $ and $B = 3\mu/\lambda$ are coefficients depending on $\lambda=(c_L \overline T_S)/\bar \ell $ and are related to fuel physical properties. From now on, we will call 'response factor' the reduced response factor defined as the real part of the transfer function $Z$:
\begin{equation}
\frac{N}{\alpha}=\Real{Z}.
\label{eq21}
\end{equation}

\section{Results and discussion}\label{sec4}
In this section, all the calculations and curves are performed with the fuel thermodynamic coefficients $A=10$ and $B=100$, corresponding approximately to orders of magnitude of values encountered in the classical fuels \citep{Prudhomme10}. Thus, relatively to the heat transfer coefficient $h$ that controls the feeding regime, and to the process characteristic times as defined in \S\ref{sec2}, and again to the influence of thermodynamic coefficients $A$ and $B$, the mean droplet mass response factor will be analysed. Figure \ref{fig2} shows response factor curves as functions of the reduced frequency $u = 3\omega\bar \tau _v $ for arbitrary values of the exchange ratio $\theta  = \bar \tau _v /\tilde \tau _T $. The five columns of diagrams correspond respectively to five values of the heat transfer coefficient: $h = 0;\;\;0.1;\;\;1;\;\;10$ and $+ \infty $.
\subsection{Effects of the heat transfer coefficient $h$}
\begin{figure}[!h]
  \centerline{\includegraphics{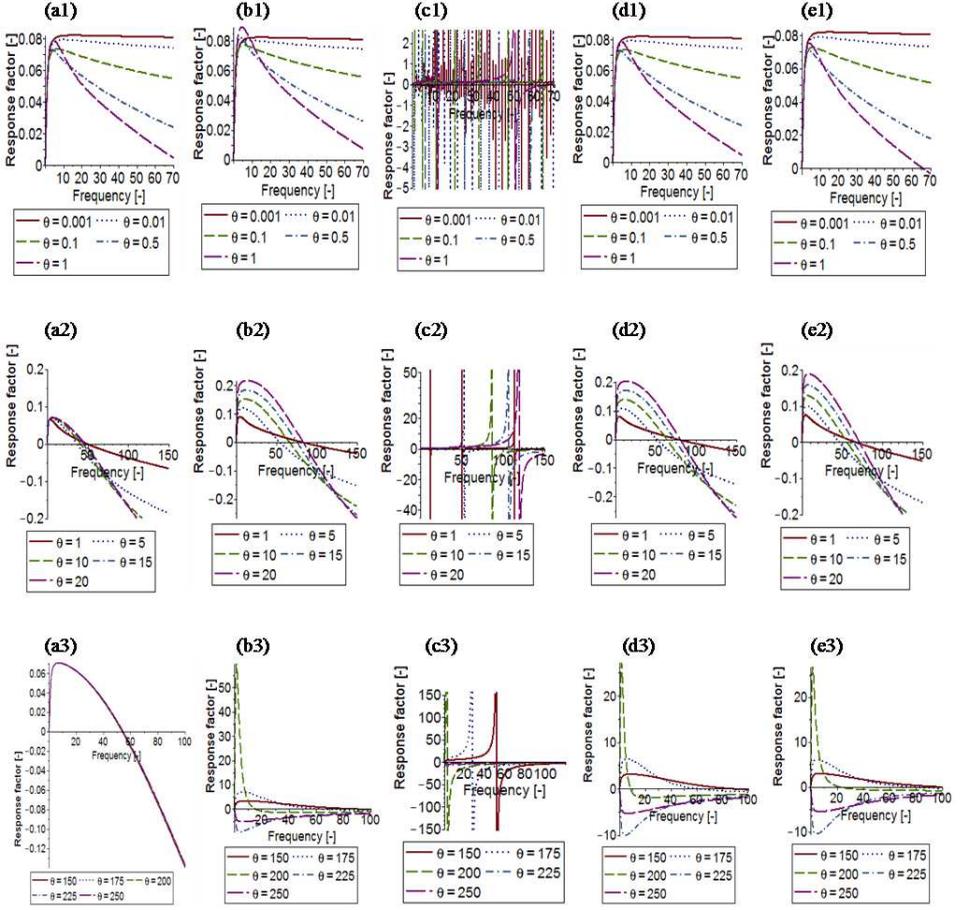}}
  \caption{Influence of heat transfer coefficient $h$ on the reduced response factor $N/\alpha  = \Real [Z(u,\theta ,h)]$ for different values of the exchange ratio $\theta$ in the mean spherical droplet model with $A = 10$ and $B = 100$. (a1), (a2) and (a3) for $h=0$ or adiabatic centre. (b1), (b2) and (b3) for $h=0.1$. (c1), (c2) and (c3) for $h=1$. (d1), (d2) and (d3) for $h=10$. (e1), (e2) and (e3) for $h=+ \infty$ or isothermal centre.}
\label{fig2}
\end{figure}

First, for $h=0$ (figures \ref{fig2}(a1), \ref{fig2}(a2) and \ref{fig2}(a3)) and for $h=\infty$ (figures \ref{fig2}(e1), \ref{fig2}(e2) and \ref{fig2}(e3)), the response factor curves seem respectively like those of the adiabatic and of the isothermal injection regimes discussed in \citet{Anani18}. In fact, these curves are identical since, for a given value of the exchange ratio $\theta$, calculations show that
\begin{equation}
E({u,\theta ,h}) \to \frac{\bar s_0 \bar r_S \theta \sin (\bar s_0 \bar r_S ) + 2\theta \cos (\bar s_0 \bar r_S ) - 3\cos (\bar s_0 \bar r_S ) - 2\theta  + 3}{\theta ( 1 - \cos( \bar s_0 \bar r_S ) )}= E( {u,\theta ,0}),
\label{eq22}
\end{equation}

when $h \to 0$, while, 
\begin{equation}
E( {u,\theta ,h}) \to -\frac{1}{2}\frac{2\theta s_0 \bar r_S \sin (s_0 \bar r_S ) + 5\theta \cos (s_0 \bar r_S ) - 3\cos (s_0 \bar r_S ) - 5\theta  + 3}{\theta ( 1 - \cos (s_0\bar r_S))} = E({u,\theta , + \infty }),
\label{eq23}
\end{equation} 
as $h \to + \infty$. The function $E( {u,\theta ,h})$ is defined by the equation (\ref{eq16}) and mentioned in the equation (\ref{eq20}). The above expressions of $E( {u,\theta ,0})$ and $E( {u,\theta ,+ \infty})$ exactly correspond to those used in the calculation of the complex transfer function $Z$ as found respectively for adiabatic and isothermal feeding regimes in \citet{Anani18}. Hence, all the comparative results highlighted concerning the two injection extreme cases in this latter reference are still valid for the present analysis. 

Secondly, according to figures 2(c1), 2(c2) and 2(c3), the response factor curves show intriguing fluctuations in their profiles when the heat transfer coefficient $h$ is fixed at 1. In that case, when the exchange ratio $\theta$ is chosen lesser than 1, the oscillations become straight chaotic although they appear relatively reduced in amplitude compared with the cases where $\theta$ is much greater than 1. Indeed, keeping $h=1$ and increasing the value of the exchange ratio $\theta$ beyond 1 until a certain threshold value to be hereafter specified, a response factor line exhibits some hyperbolic pattern with high peaks value along the reduced frequency axis as in figures \ref{fig2}(c2) and \ref{fig2}(c3). Moreover, once the heat transfer coefficient slightly differs from 1, the curves tend to show more lower fluctuations in their profiles even if $h$ remains very close to 1 as for $h=0.95$ or $h=1.05$, as many examples not illustrated with figures. For comparison, a unity value of a heat transfer coefficient may characterise a radiation heat transfer processing from the flame to the chamber wall. According to \citet{Santos08} for example, the radiative power is highly nonlinear and varies at the first order as the fourth power of the local instantaneous temperature. It may be admitted that, even in fuel injection processes, this specific value of the liquid-liquid heat transfer coefficient ($h=1$) can strongly influence on the evaporating mass release response in a perturbed environment. At this point, experimental investigations are necessary for further clarifications. 

\subsection{Effects of process characteristic times}
The vaporization response of a LOX droplet to oscillatory ambient conditions has been computed over a wide range of frequencies and the results were applied to prototypical cases pertinent to liquid rocket combustion instabilities \citep{Sirignano94}. It has been shown that the peak frequency for the computed response factor is correlated to the droplet lifetime. Indeed, as also recorded in \citet{Anani18}, the peak value of a response factor curve, whenever it exists, occurs at the same peak reduced frequency $u_p$ about 3 (see among others figures \ref{fig2}(b1), \ref{fig2}(b2) for $h=0.1$ or \ref{fig2}(d1), \ref{fig2}(d2) for $h=10$). In the mixed feeding ($h>0$) as well as in the extreme cases of adiabatic and isothermal injection regimes, one has $u_p  = 3\omega \bar \tau _v  \approx  3$. This relation implies $\bar \tau _v  \approx 1/\omega$, meaning that the injected liquid residence time $\bar \tau _v$ is at the same order of magnitude as the oscillation period $1/\omega$. Now, the mean residence time $\bar \tau _v$ of a continuously fed droplet can be equated to the mean lifetime of a free droplet in the spray. Thus, whenever positive responses appear in the system, regardless of the value of the heat transfer coefficient, the vaporization rate can fully respond to the acoustic oscillations, only when the mean droplet lifetime equals the period of ambient pressure oscillations.
\subsection{Influence of the thermodynamic coefficient $B$}\label{ssec43}
As they occur about a fixed value of the thermal exchange ratio ($\theta  \approx 200$), the sharp changes noted in the response factor curve profiles are not related to some particular values of the heat transfer coefficient, but rather to a specific value of $\theta$. As in \citet{Anani18}, those rapid changes in curve profiles around the reduced frequency $u_p  \approx 3$ can be proved depending on a specific value of $\theta$ in relation to the liquid fuel thermodynamic coefficient $B = 3\mu /\lambda $. In order to determine the threshold value $\theta_d$ of the thermal exchange ratio at which abrupt changes intervene in the curve profiles, the ratio $x = u/\theta  = \omega \bar \tau _T /3$ may be particularly useful. Indeed, the thermal diffusion time $\bar \tau _T $ and the frequency of the oscillating wave $\omega $ do intervene in this ratio but not the residence time $\bar \tau _v$. This ratio can then be taken negligible at the fixed peak frequency $u_p  = 3\omega _p \bar \tau _{vp}  \approx 3$ provided that the thermal transfer time by diffusion $\bar \tau _T$ is taken negligible, compared to the oscillation period $1/\omega _p $ or to the residence time $\bar \tau _{vp}$ as $1/\omega _p\approx\bar \tau _{vp}$ at $u_p$. Therefore, whenever $h > 0$, the second-order truncated expansion of the complex transfer function $Z(u,\theta ,h)$ in the neighbourhood of $x=0$ while assuming $u$ closer to $u_p$, gives the expression:
\begin{equation}
Z(u,\theta ,h) \approx \frac{\mathrm{i}u\left( A +\frac{\theta}{2}-\frac{3}{2}\right) }{(1 + \mathrm{i}u)\left( B-\frac{\theta}{2}+\frac{3}{2}\right) }
\label{eq24}
\end{equation} 
which is no more dependent on the heat transfer coefficient $h$. But when $h=0$ i.e. in the adiabatic centre regime, the computations lead to the following approximation: $Z(u,\theta ,0) \approx {\mathrm{i}u(A -3)}/[{(1 + \mathrm{i}u)(B+3)}]$. In consequence, once a not null heat transfer coefficient is introduced in the feeding process, the value of $\theta$ around which mass response factor curves exhibit a sharp top at the peak frequency $u_p$, can be deduced from the expression (\ref{eq24}) by equating the denominator to 0. Thus, $\theta _d  = 2B + 3 = 203$ for $B=100$. Moreover, once $\theta$ becomes greater than $\theta_d$, using again the approximation (\ref{eq24}), one has $\Real(Z)\leq 0$ whenever $h>0$, and the corresponding response factor curves show only negative response for all frequencies as shown in figures \ref{fig2}(b3) for $h=0.1$, \ref{fig2}(c3) for $h=1$ and \ref{fig2}(d3) for $h=10$. As in \citet{Hsiao11} and in \citet{Ren19}, many publications highlight the rapid changes noticed in fluid thermophysical properties in connection with critical and supercritical vaporization processes as contributing to abrupt changes in mass release response. However, as shown by the present study, an abrupt or a completely damped vaporization frequency response may occur during subcritical combustion processes provided that certain specific boundary conditions are imposed.
\section{Conclusions}
Through the introduction of a heat transfer coefficient in the liquid fuel injection process, this study has extended to a more generalized feeding regime the results on the vaporization frequency response to ambient pressure oscillations. An idealized configuration of the mean droplet has permitted to compute the mass frequency response of the vaporizing spray. The effects of the heat transfer coefficient and of the characteristic times, and again of the thermal exchange ratio are found effective for driven or dampen instabilities. It was shown that, whenever positive responses appear in the system, the peak value is reached at a particular frequency, where the residence time of the mean droplet matches the period of the ambient pressure oscillations. Except for the case where the heat transfer coefficient value is equal to one, response factor curves exhibit a single abrupt peak response at the particular frequency as the thermal exchange ratio approaches a certain value. The latter is shown equal to a simple affine function of a thermodynamic coefficient related to fuel physical properties. Once this threshold value of the thermal exchange ratio is passed over, the related factor curve shows only negative response for all frequencies even if the heat transfer coefficient value is maintained at one. The results are also found similar to those previously obtained in the adiabatic and isothermal feeding regimes. Indeed, mass response factors in such extreme cases of fuel injection are recovered as simple limit points. The above-mentioned results may be beneficial for instability control in combustion processes.

\bibliographystyle{jfm}

\bibliography{frequency}

\begin{thebibliography}{14}
\expandafter\ifx\csname natexlab\endcsname\relax\def\natexlab#1{#1}\fi
\def\au#1{#1} \def\ed#1{#1} \def\yr#1{#1}\def\at#1{#1}\def\jt#1{\textit{#1}}
  \def\bt#1{#1}\def\bvol#1{\textbf{#1}} \def\vol#1{#1} \def\pg#1{#1}
  \def\publ#1{#1}\def\arxiv#1{#1}\def\org#1{#1}\def\st#1{\textit{#1}}

\bibitem[Abramzon \& Sirignano(1989)]{Abramzon89}
{\sc Abramzon, B. \& Sirignano, W.~A.} \yr{1989} \at{Droplet vaporization model for spray combustion 
 calculations.}  \jt{Int. J.~Heat Mass Tran.} \bvol{32}, \pg{1605--1618}.
 
\bibitem[Anani {\em et al.}(2018)]{Anani18}
{\sc Anani, K., Prud'homme, R., \& Hounkonnou M.~N.} \yr{2018} \at{Dynamic response of a vaporizing 
 spray to pressure oscillations: Approximate analytical solutions.} \jt{ 
 Combust. Flame} \bvol{193}, \pg{295--305}.

\bibitem[Candel {\em et al.}(2013)]{Candel13}
{\sc Candel, S., Durox, D., Schuller, T., Darabiha, N., Hakim, L. \& Schmitt T.} \yr{2013} \at{Advances
 in combustion and propulsion applications.} \jt{ 
 Eur. J.~Mech. B-Fluid} \bvol{40}, \pg{87--106}.

\bibitem[Heidmann \& Wieber(1966)]{Heidmann66}
{\sc Heidmann, M.~F. \& Wieber, P.~R.} \yr{1966} \bt{Analysis of frequency response characteristics of 
 propellant vaporization}. {\em  Tech. Rep.\/} X-52195. \org{NASA Tech.\ Mem.}

\bibitem[Hsiao {\em et al.}(2011)]{Hsiao11}
{\sc Hsiao, G.~C., Meng, H., \& Yang V.} \yr{2011} \at{Pressure-coupled vaporization response 
 of n-pentane fuel droplet at subcritical and supercritical conditions.} \jt{
 Proc.\ Combust.\ Inst.} \bvol{33}, \pg{1997--2003}.

\bibitem[Nair \& Sujith(2014)]{Nair14}
{\sc Nair, V. \& Sujith, R.~I.} \yr{2014} \at{Multifractality in combustion noise: 
 predicting an impending instability.} \jt{J.~Fluid Mech.} \bvol{747}, \pg{635--655}.

\bibitem[Prud'homme {\em et al.}(2010)]{Prudhomme10}
{\sc Prud’homme, R., Habiballah, M., Matuszewski, L., Mauriot, Y. \& Nicole A.} \yr{2010} \at{Theoretical
 analysis of dynamic response of a vaporizing droplet to a acoustic oscillation.} \jt{
 J.\ Propul.\ Power} \bvol{26}, \pg{74--83}.

\bibitem[Ren {\em et al.}(2019)]{Ren19}
{\sc Ren, J., Marxen, O. \& Pecnik, R.} \yr{2019} \at{Boundary-layer stability of supercritical 
  fluids in the vicinity of the Widom line.} \jt{J.~Fluid Mech.} \bvol{871}, \pg{831--864}.

\bibitem[Santos {\em et al.}(2008)]{Santos08}
{\sc Santos, R. G.~D., Lecanu, M., Ducruix, S., Gicquel, O., Iacona, E. \& Veynante D.} \yr{2008} \at{Coupled 
 large eddy simulations of turbulent combustion and radiative heat transfer.} \jt{
 Combust. Flame} \bvol{152}, \pg{387--400}.
 
\bibitem[Sirignano {\em et al.}(1994)]{Sirignano94}
{\sc Sirignano, W.~A., Delplanque, J.-P., Chiang, C.~H. \& Bhatia R.} \yr{1994}
 \at{{Liquid-propellant droplet vaporization: a rate controlling process for 
 combustion instability}}. \bt{In {\em Liquid rocket engine combustion instability\/}
 (ed. \ed{V. Yang \& W.~E. Anderson})}, \pg{pp. 307--343}. \publ{Reston}.

\bibitem[Slavyanov \& Lay(2000)]{Slavyanov00}
{\sc Slavyanov, S.~Y. \& Lay W.} \yr{2000} \at{{The Heun class of equations}}. \bt{In {\em 
 Special Functions: A Unified Theory Based on Singularities\/}},
 \pg{pp. 97--162}. \publ{Oxford University Press}.
 
\end{thebibliography}

\end{document}